\documentclass[preprint]{article}

 \usepackage{neurips_2025}

\usepackage[utf8]{inputenc} 
\usepackage[T1]{fontenc}    
\usepackage{hyperref}       
\usepackage{url}            
\usepackage{booktabs}       
\usepackage{amsfonts}       
\usepackage{nicefrac}       
\usepackage{microtype}      
\usepackage{xcolor}         
\usepackage{float}
\usepackage{algorithm}
\usepackage{amsmath}
\usepackage[noend]{algpseudocode}
\usepackage{graphicx, caption, subcaption}
\newtheorem{theorem}{Theorem}[section]

\newtheorem{lemma}[theorem]{Lemma}

\newtheorem{remark}[theorem]{Remark}
\newenvironment{proof}{\paragraph{Proof:}}{\hfill$\square$}
\newcommand{\RR}{\mathbb{R}}

\newcommand{\PR}{\mathbb{P}}

\title{A Goemans-Williamson type algorithm for identifying subcohorts in clinical trials}

\author{%
Pratik Worah\thanks{pworah@uchicago.edu (Work done while the author was visiting NYU. Author's current affiliation: Google Research).}
}

\begin{document}

\maketitle

\begin{abstract}
We design an efficient algorithm that outputs tests for identifying predominantly homogeneous subcohorts of patients from large in-homogeneous datasets. Our theoretical contribution is a rounding technique, similar to that of Goemans and Wiliamson (1995), that approximates the optimal solution within a factor of $0.82$. As an application, we use our algorithm to trade-off sensitivity for specificity to systematically identify clinically interesting homogeneous subcohorts of patients in the RNA microarray dataset for breast cancer from Curtis et al. (2012). One such clinically interesting subcohort suggests a link between LXR over-expression and BRCA2 and MSH6 methylation levels for patients in that subcohort.
\end{abstract}
\section{Introduction}
A linear separator in $\RR^d$ is a basic tool used in statistics and machine learning to partition a set of points into two subsets. 
Given a set of Red and Blue points to be separated, there may not exist a linear separator that separates most of the Red and Blue points. Moreover, even if one exists, then it can be expected to have most coefficients to be far from zero, i.e., it's not sparse, unless there are structural assumptions on the dataset. On the other hand, if we want a linear separator that separates a small (but non-trivial) percentage of Red points from Blue points, then a sparse linear separator may exist. Such sparse separators also have the advantage that they are likely to be interpretable. Therefore, the focus  shifts to identifying a subset of Red points to carve out from the rest, so that: (1) the number of Blue points in the identified subset is relatively small (specificity is high); (2) the relative number of Red points in the subset, compared to the total number of Red points, is large (sensitivity is high); and (3) the number of non-zero coefficients in the separator is small (the separator is sparse). 

In this paper, we devise a semidefinite programming (SDP) based algorithm (see Algorithm~\ref{algmain}) for computing sparse linear separators that identify homogeneous subgroups. We prove theoretical guarantees on the specificity and sensitivity of the separator (see Theorems~\ref{thm:main} and~\ref{thm:spec}). We apply our algorithm to design a test for identifying subgroups of metastatic breast cancer cases from a large breast cancer dataset~\cite{metabric, metabric2, metabric3}. We empirically study the trade-offs between the sensitivity, specificity and sparsity of the tests output by our algorithm in Section~\ref{sbs:pred}. Finally, in Section~\ref{sbs:nn}, we use our algorithm to identify clinically interesting subgroups of breast cancer patients.

\subsection{Related literature}

Perhaps the most well-known subgroup identification algorithm is the Patient Rule Induction Method (PRIM)~\cite{ff} and its variants~\cite{rlbk,prim-cart}. These are local search algorithms with heuristics designed to gradually trade-off sensitivity for specificity. At it's crux, the better-known PRIM algorithm relies on identifying one or more hyper-rectangles, where points in the interior of a hyper-rectangle constitute the identified subgroup. While such bump hunting heuristics are hard to analyze theoretically, explainable clustering algorithms using axis-aligned thresholding cuts have been theoretically analyzed (for example~\cite{SIELING2008394},~\cite{expl-kmeans1} and~\cite{expl-kmeans2}). There are at least two significant differences between algorithms using axis aligned hyper-rectangles (equivalent to thresholding rules on coordinates) and our approach. First, such rule-based algorithms are almost always restricted to axis-aligned thresholding cuts, as opposed to sparse cuts (as in this paper); even though the latter are also intuitively interpretable. Second, each axis aligned half-space can also be thought of as node in a decision tree, so computing subgroups that are identified by one or more axis-aligned hyper-rectangles corresponds to computing a decision tree, where the underlying objective is chosen to balance sensitivity and specificity. It is well-known that computing optimal decision trees for most reasonable objectives is not just a NP-hard problem, but most versions are even hard to approximate within any constant factor (see for e.g.~\cite{gupta1, koch2024superconstant}). Intuitively, this is because of close connections between decision trees and circuit complexity. In contrast, in this paper, we translate subgroup identification with sparse linear cuts to a MAX-CUT like problem (cf.~\cite{gw}), which admits constant factor approximation guarantees via convex optimization (an approximation factor of $0.82$ in our case). Moreover, there are practical examples where a sparse linear hyperplane, as opposed to a rule based algorithm, is used to identify interesting subcohorts. For example, the papers~\cite{fbtest} and~\cite{meld} describe diagnostic scores, essentially sparse regression models, that are widely used in gastroenterology.

\section{Overview of our results}
{\em Motivation:} Suppose we are given a set of patients in different disease states. For example, in case of cancer patients, some of them may have metastasis but others may not. We want to identify a small set of genes or proteins (or some other markers) that allows us to test and select patients that are in the same state, i.e., we want to design tests that identify predominantly homogeneous subcohorts (subsets) of breast cancer patients with metastasis. For example, if roughly 10\% of the cancer patients have metastasis, then we want to design an algorithm that uses a small number of gene expressions to identify a large enough subcohort, say 2\% of the overall patient population, of which 70\% have metastasis.

Clearly, there are trade-offs involved. If we want the identified subcohort of patients to be a larger fraction of the overall patient population, i.e., if we want the {\em sensitivity} to be large, then the probability that most patients in the subcohort will have metastasis, will be smaller, i.e., the {\em specificity} will be smaller (equivalently the subcohort will be more inhomogeneous). Furthermore, if we want the set of identifying markers to be small, i.e., if we want the designed test to be {\em sparse}, then the specificity is going to be lower (assuming sensitivity is kept constant). On the other hand, if we want to have sparse tests and high specificity in our subcohort, then the sensitivity (which corresponds to subcohort size) will be reduced. 

{\em Contribution:} In this paper, we formalize and solve the subcohort identification problem using Algorithm~\ref{algmain}, 
which is a generalization of the Goemans-Williamson semidefinite programming (SDP) algorithm for solving MAX-CUT~\cite{gw}. Our theoretical contribution can be summarized as follows:
\begin{itemize}
    \item We prove that the ratio of the size of subcohorts that are identified by Algorithm~\ref{algmain} vs. the optimal subcohorts is lower bounded by an absolute constant $\alpha\simeq 0.82$ (see Section~\ref{sbs:theory} for details). This implies bounds on the sensitivity and specificity as well (Theorems~\ref{thm:main} and~\ref{thm:spec}).
\end{itemize}
The theoretical result should be independently interesting, as it generalizes the Goemans-Williamson calculations. While similar SDP rounding algorithms for similar problems have been analyzed before (for e.g.~\cite{feige1,rpr2} and~\cite{potechin}), we are not aware of any similar analysis for the problem in our paper.\footnote{In a nutshell, the crux of the Goemans-Williamson guarantee relies on computing the maximum ratio of the chord length to the minor arc length in a unit circle, while in our case it hinges on bounding the ratio of the "perimeter" of a triangle to the perimeter of the circumcircle, the latter is assumed to be of unit radius. It is surprising that a Goemans-Williamson SDP rounding works here, as our objective function is not sign definite. Usually more complicated rounding techniques, like~\cite{alon-naor} are required in such cases.}

Furthermore, we empirically study the trade-offs mentioned above, for Algorithm~\ref{algmain}, in the context of identifying subcohorts of metastatic breast cancer using the METABRIC dataset~\cite{metabric,metabric2,metabric3}.\footnote{The dataset contains RNA microarray data of approximately 2,000 breast cancer patients, and methylation data for approximately 13K genes in about 1,500 breast cancer patients. About a thousand patients had one or more positive lymph nodes upon biopsy (metastasis), while the remaining did not (non-metastasis).} 

For the empirical part, our first result identifies subcohorts of predominantly metastatic breast cancer cases using sparse linear tests based on expression of nuclear receptors, i.e., we want the ratio of the metastatic cases to all breast cancer cases to be high for the identified subcohort. We denote this ratio as the {\em specificity} of our test. The number of metastatic cases in the identified subcohort as a percentage of all breast cancer cases is denoted as the {\em sensitivity} of the test. The number of genes used is roughly proportional to the sparsity parameter of Algorithm~\ref{algmain}. Our first empirical result relates the specificity, sensitivity and the number of genes used, as follows: 
\begin{itemize}
  \item We show that specificity and sensitivity both increase with the number of genes used in the test (see Figure~\ref{fig:comp1}(a) and Subsection~\ref{sbs:pred} for details). In particular, we obtained subcohorts of size $\sim 20$\% of the subsampled patient population, with about 57\% specificity, when we use roughly 15 out of 48 nuclear receptors in the designed test.
  
  \item Moreover, if we define our subcohort as an intersection of half-spaces from two different runs of Algorithm~\ref{algmain} (over two subsamples of the dataset), then we can increase the specificity by sacrificing on sensitivity (see Figure~\ref{fig:comp1}(b)). In particular, we obtained subcohorts of size $\sim 7$\% of the subsampled patient population, with about 67\% specificity, when we use roughly 45 out of 48 nuclear receptors in the test.
\end{itemize}
Our second empirical result, is an application of Algorithm~\ref{algmain} to identify clinically interesting large homogeneous subcohorts of patients. 
One example of a clinically interesting problem is to identify statistically significant associations between changes in methylation percentages and changes in nuclear receptor expressions. Methylation of tumor suppressor genes is known to be a potential cause of progression of breast cancers~\cite{brcgenes}, while nuclear receptors are often used as targets of many drugs~\cite{nrdrugs}. Therefore, we are attempting to associate the likely cause of the pathology to a likely  treatment target, for a specific subcohort of patients.

The question arises: why is identifying such associations difficult or novel, after all, it involves only standard deviation computation? A naive average over the entire cohort of patient population, where one indiscriminately averages over non-homogeneous data, may make confidence intervals larger than the separation between the means. Thus deviations from the mean are no longer statistically significant. This is exactly what happens in Figure~\ref{fig:poplevel}, where we compute the average expression of nuclear receptors and methylation percentages in metastatic and non-metastatic breast cancer patients, over the entire cohort of about 2,000 patients. Based on the figure, one can not differentiate between metastatic vs non-metastatic breast cancer using deviations and simple Z-scores of any of the 90 odd genes plotted -- the population is inhomogeneous -- confidence intervals are too wide to differentiate between the two classes of patients with statistical significance. Similarly, if one computes averages over many small subsets of the population, across subcohorts, then the confidence intervals are likely to be too large due to small sample size. Therefore, to deduce clinically interesting conclusions, we need to identify and compute the deviations over homogeneous subcohorts. 

In Subsection~\ref{sbs:nn}, we trade-off sensitivity for specificity using Algorithm~\ref{algmain} to make the identified subcohorts more homogeneous. One clinically interesting subcohort is shown in Figure~\ref{fig:assoc}. For this subcohort, we observe:
\begin{itemize} 
    \item There is significant methylation increase in genes BRCA2 and MSH6, and nuclear receptor NR1H2 (equivalently LXRB) is significantly over-expressed, both relative to the baseline population. Therefore, one may hypothesize that methylation of the former leads to downstream over-expression of the latter.  
    \item Furthermore, significant over-expression of LXRB receptor gene suggests that patients in the subcohort may be good candidates for LXR-inverse agonists, if supported by clinical evidence. 
    
    Here, it is worthwhile to note that both LXR-inverse agonists~\cite{lxr-ant} and LXR agonists~\cite{lxrbeta} have been proposed to be beneficial for various subsets of breast cancer patients. Our algorithm, and its output test, may offer a principled way to identify genomic pathways underlying the disease. Moreover, it may also help identify patients, who could benefit from use of LXR-inverse agonists or LXR agonists (or neither).
\end{itemize}
Currently, clinically homogeneous subgroups of breast cancer patients are often identified by single genes. For example, ER+ patients, i.e., those breast cancer patients with ER nuclear receptor expression above a certain threshold, are treated with a certain protocol and vice versa~\cite{bc-protocol}. This is a likely suboptimal determination of patient subcohorts and resulting treatment protocols. In contrast, our algorithm presents a systematic approach with a potentially better handle on sensitivity, specificity and sparsity trade-offs.



{\em Organization:} We begin with formal algorithmic description and theoretical details in Section~\ref{sbs:theory}, which is followed by empirical results on the breast cancer dataset in Section~\ref{sbs:genomics}. 

\section{Underlying theoretical problem}\label{sbs:theory}
We begin with an abstract description of our problem. We are given a complete bipartite graph $G\equiv (U,V)$, where the vertices are embedded in a high dimensional space $\RR^d$. Our goal is to compute a hyperplane $(h,\theta)$, i.e., $\langle h, x\rangle \le -\theta$,\footnote{Here $x\in\RR^d$ is the input specified embedding of a given vertex in $G$ into $\RR^d$.} with few non-zero coefficients in $h$, that separates a subset $S\equiv (S_U,S_V)\subset U\cup V$ of vertices from the rest of the vertices in $G$. In addition, we want $S_U$ to be much larger than $S_V$, and $S$ to be large. 

It is easy to see that the computed subset $S$ satisfying the above requirements is precisely the subcohort meeting the requirements from the previous discussion. 

Without loss of generality, we may assume that the input also specifies a vertex $u_0\in S_U$, since one can always iterate over all vertices in $U$. Note that, $S$ is large and $|S_U|\gg|S_V|$, together imply that the cut $(S_U,V\setminus S)$ is large. Therefore, one may formulate the above as a MAX-CUT type integer program, where one associates a $\pm 1$ variable $z_u$ with each of the $n$ vertices in $G$, and wants to maximize the objective:
\begin{equation}\label{eqn:obj1}
   \max_{h\in\RR^d,\atop{\vec{z}\in\{\pm 1\}^n}} \sum_{u\in U,\atop{u'\in V}}\frac{(z_u-z_{u'})^2+(z_{u_0}-z_{u'})^2-(z_u-z_{u_0})^2}{3}.
\end{equation}
Maximizing the size of such a cut is equivalent to maximizing the size of $S$, while making $|S_U|$ larger than $|S_V|$. Without any further constraints, the trivial solution $S_U\equiv U,  S_V\equiv V$ is optimum. However, the separation and sparsity constraints make the optimum non-trivial. These constraints may be enumerated as follows.
\begin{itemize}
    \item Separating edges between $U\setminus \{u_0\}$ and $V$: These edges should only contribute to the determination of the separating hyperplane if $u\in U\setminus \{u_0\}$ is not cut, i.e., $u$ and $u_0$ are on the same side ($u$ is in the subcohort). Hence, for every such $u\in U\setminus\{u_0\}$, we add the constraint:
    \begin{equation}\label{eqn:sep1}
        \langle h, x_u\rangle \le -\theta z_uz_{u_0}
    \end{equation}
    \item Separating edges between $u_0$ and $V$: These edges should only contribute to the determination of the separating hyperplane if $u'\in V$ is cut, i.e., $u'$ and $u_0$ are on different sides ($u'$ is not in the subcohort). Hence, for every such $u'\in V$, we add the constraint:
    \begin{equation}\label{eqn:sep2}
        \langle h, x_{u'}\rangle \ge \theta (1-z_{u'}z_{u_0})
    \end{equation}
    \item Sparsity: We want the designed test to use few genes, i.e., we want its coefficients to be sparse. Hence we bound the $\ell_1$ norm of its coefficients (see for example~\cite{rauhut}) by an input parameter $s$, which is expressed as: 
    \begin{equation}\label{eqn:sparsity-param}
        \|h\|_1\le s.
    \end{equation}
\end{itemize}

\begin{remark}
It is worth noting that the formulated problem above is very similar to maximum margin classification using linear support vector machines. Thus, it is possible to generalize it to non-classifiers using the "kernel trick" (see for example~\cite{vapnik}). The only exception being the sparsity constraint above, which needs to be adapted for the chosen kernel. It may be possible to generalize the sparsity constraint for low degree polynomial kernels using standard compressed sensing methods (see for example~\cite{rauhut}). However, such non-linear extensions are not the focus of this work.
\end{remark}

One may relax the discrete objective function and the linear constraints on the embedding into a SDP, as specified below. Note that the SDP relaxation can be efficiently solved using a standard solver unlike the discrete objective function.

\begin{eqnarray}\label{eqn:sdp}
&{\max_{h\in\RR^d,\atop{\forall i:\ \|v_i\|_2=1}}} &{\sum_{u\in U,\atop{u'\in V}}\frac{2-2\langle v_u,v_{u'}\rangle-2\langle v_{u_0},v_{u'}\rangle+2\langle v_u,v_{u_0}\rangle}{3},}\\
{ s.t.}&{\forall u\in U   :}&\ {\langle h,x_u\rangle\le -\theta \langle v_u,v_{u_0}\rangle}\label{eqn:rlx1}\\
&{\forall u'\in V   :}&\ {\scriptstyle\langle h,x_{u'}\rangle \ge \theta(1-\langle v_{u'},v_{u_0}\rangle)}\label{eqn:rlx2}\\
&&{\|h\|_1\le s,}
\end{eqnarray}

where $x_u$ are the embeddings in $\RR^d$, $s$ is the sparsity parameter, and $\theta$ is the separation threshold parameter, all to be specified with the input. Algorithm~\ref{algmain} essentially maps the vector solution of the SDP to $\pm 1$ values, to obtain the subcohort (corresponding to $\{i\in U\cup V:\ z_i=1\}$). 

\begin{algorithm}[htb]
\caption{Identify subcohort}\label{algmain}
\begin{algorithmic}[1]
\Statex {\bf Input:} Two sets of vertices $(U,V)$ labeled $0$ and $1$ respectively, embedded in $\RR^d$; vertex $u_0\in U$, sparsity parameter $s\in(0,\infty)$, and separation threshold $\theta\in\RR$.
\Statex {\bf Output:} If feasible, compute a hyperplane $h\in\RR^d$ and a subcohort $S\equiv (S_U,S_V)\subset U\cup V$ such that: (1) $u_0\in S_U$, (2) $\forall u\in S_U:\ \langle x_u, h\rangle\le \theta$ and $\forall v\in S_V:\ \langle x_v, h\rangle\ge 0$, (3) $\|h\|_1\le s$, and (4) $|S_U|\gg|S_V|$. Note: Here $x_u$ and $x_v$ are the $\RR^d$ embeddings corresponding to vertices $u$ and $v$ (from the input) and $\|h\|_1$ denotes the $\ell_1$ norm of the coefficients of $h$.
\Statex
\State Solve the SDP relaxation (Equation~\ref{eqn:sdp}) to obtain the solution vectors $\{v_i\}_{i\in[n]}$.
\State If SDP infeasible, {\bf return} infeasible.
\State $\forall i\in\{1,...,n\}:\ z_i\leftarrow\mathrm{sign}(\langle v_i,\gamma_i\rangle)$, where $\gamma_i\sim\mathcal{N}(0,1)$.
\State \bf{return} $\vec{z}, h$
\end{algorithmic}
\end{algorithm}

\noindent In order to show that the sensitivity and specificity of the test output by the SDP solution above are not too far from the optimum, we prove the following approximation guarantee (see the appendix, at the end, for proofs).
\begin{theorem}[Sensitivity]\label{thm:main}
Let $S^*=(S^*_U,S^*_V)$ denote the optimal subcohort such that $|S_U^*|\ll |U|$, then Algorithm~\ref{algmain} computes a subcohort $S=(S_U,S_V)$, such that $|S_U|\ge \alpha|S^*_U|$, where the constant $\alpha\simeq 0.82$. Equivalently, $\frac{|S_U|}{|U|+|V|}\ge \alpha\cdot\frac{|S^*_U|}{|U|+|V|}$, i.e., the sensitivity of the output test is within $0.82$ of the optimum.
\end{theorem}

\begin{theorem}[Specificity]\label{thm:spec}
For $\kappa\in(0,1)$, let $\kappa\cdot|U||V|$ denote the fraction of edges in the optimal sparse cut separating $U$ from $V$. The specificity of the test output by Algorithm~\ref{algmain} is lower bounded by the solution to the following optimization problem:
\begin{equation}\label{eqn:spec-opt}
    \min_{x,y\in(0,1)} \frac{x\cdot|U|}{(1-y)|V|+x|U|}\qquad \mathrm{s.t.}\quad x\cdot y\ge\tilde{\alpha}\cdot\kappa,
\end{equation}
where $\tilde{\alpha}=\frac{\alpha}{2}\simeq0.41$.
\end{theorem}
Note that the solution of the optimization problem in Equation~\ref{eqn:spec-opt} grows as $\Omega(\sqrt{\kappa})$, when $|U|$ is assumed equal to $|V|$ (this can be seen by substituting $y=\frac{\tilde{\alpha}\cdot\kappa}{x}$ in the objective). A plot of the solution surface, together with the proof, is given in the appendix (Section~\ref{sec:app-spec}).

While the optimum hyperplane separator $h$ remains unchanged after rounding though the input specified threshold $\theta$ is perturbed during rounding. However, we may bound the perturbation amount by using Grothendieck's identity, i.e.,
\begin{equation}
  \mathbb{E}\left[\mathrm{sign}(z_u)\mathrm{sign}(z_{u_0})\right]=\frac{2}{\pi}\arcsin\left(\langle v_u,v_{u_0}\rangle\right).
\end{equation}as follows.
\begin{theorem}\label{thm:constr}
The rounded solution from Algorithm~\ref{algmain} obeys the (slightly weaker) separation constraints:
\begin{eqnarray}
    \langle h, x_u\rangle &\le& -\frac{2\theta}{\pi} z_uz_{u_0}\\
    \langle h, x_{u'}\rangle &\ge& \theta \left(1-\frac{2}{\pi}(z_{u'}z_{u_0})\right)
\end{eqnarray}
as opposed to the corresponding constraints in Equations~\ref{eqn:sep1} and~\ref{eqn:sep2}.
\end{theorem}
\section{Experiments: Genomics applications}\label{sbs:genomics}

In this subsection, we use Algorithm~\ref{algmain} to generate tests\footnote{A test is just the output hyperplane of Algorithm~\ref{algmain} -- a weighted average of gene expression levels -- that evaluates to a score less than the given threshold $\theta$ for metastatic cases, and above $\theta$ for non-metastatic cases.} that identifies subcohorts with predominantly metastatic cases in a breast cancer dataset. Our empirical results may be divided into two parts:
\begin{itemize}
    \item In Subsection~\ref{sbs:pred}, we examine the quality of the identified subcohorts. A random test, i.e., a random half-space, can be expected to generate subcohorts with less than 50\% metastasis cases (as the number of non-metastasis patients is higher in the dataset). In Figure~\ref{fig:comp1}(a), we show that the linear tests generated by Algorithm~\ref{algmain} identify subcohorts where greater than 50\% of the cases are metastatic (thus outperforming random). Moreover, if instead of using one hyperplane to identify a subcohort, we use two hyperplanes to identify a subcohort then the identified subcohorts have 60-70\% of their cases as metastatic (see Figure~\ref{fig:comp1}(b)). We also compare our algorithm's specificity (for a given sensitivity) with a version of the well-known PRIM algorithm for subgroup discovery (this comparison is in the Appendix Section~\ref{sec:prim}).
    \item In Subsection~\ref{sbs:nn}, we use Algorithm~\ref{algmain} to identify a clinically interesting subcohort of breast cancer patients -- those that may respond to LXR inverse agonists. In the process, we also associate epigenetic changes with changes in nuclear receptor expressions. It is known~\cite{brcgenes}, that about $40$ genes act as tumor suppressors, and their methylation leads to reduced expression and potentially cancer progression. Moreover, in~\cite{nrdrugs,nrdrugs2}, the authors discuss the role of nuclear receptors as drug targets. Therefore, it is natural to look for associations between significant changes in nuclear receptor expressions and significant changes in methylation levels of breast cancer related genes (over population baseline levels) -- the latter can guide treatment drug targets from the former. We show that, even though at the entire dataset level such associations do not exist, but at the subcohort level they do exist (see Figure~\ref{fig:assoc}).
\end{itemize}

\noindent {\em Dataset:} The results in this section are based on the METABRIC dataset~\cite{metabric, metabric2, metabric3}, which contains RNA microarray data from biopsy samples of approximately 1904 (911 with positive lymph nodes and 993 without positive lymph nodes) breast cancer patients, and methylation data for approximately 13K genes in about 1406 (669 with positive lymph nodes and 737 without positive lymph nodes). In this paper, we (crudely) denote patients with one or more positive lymph nodes upon biopsy as metastasis cases, while the rest as non-metastasis cases.

\subsection{Predicting breast cancer metastasis}\label{sbs:pred}
Recall that, the {\em sensitivity} of Algorithm~\ref{algmain} is the percentage of metastatic cases that are correctly classified to be in the subcohort, while the {\em specificity} is the percentage of metastatic cases in the identified subcohort. The number of genes used by the test is defined as the number of coefficients of $h$ whose magnitude is at least $\frac{1}{10}$ of the maximum magnitude coefficient in $h$. Due to well known results from $\ell_1$ norm minimization~\cite{rauhut}, the number of genes used should be roughly proportional to the sparsity parameter input to the algorithm (Equation~\ref{eqn:sparsity-param}).

\begin{figure*}[htb!]
    \centering
    \subfloat[][The number of metastasis cases (true positives) and non-metastasis cases (false positives) in the identified subchort using a single test from Algorithm~\ref{algmain}. Sensitivity is proportional to the height of the blue bar, while Specificity is proportional to the ratio between the height of blue and red bars. The sparsity parameter (x-axis) is  proportional to the number of genes used.]{\includegraphics[scale=0.4]{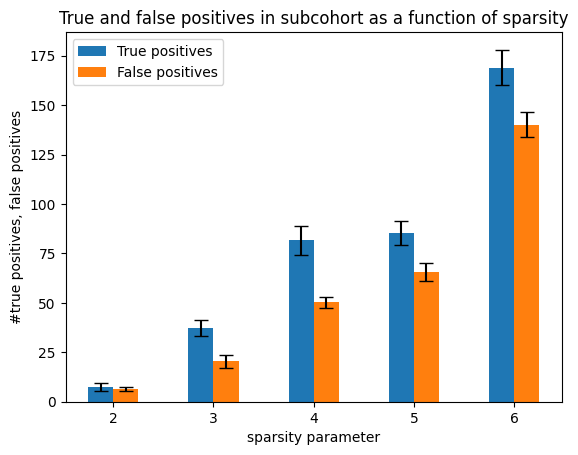}}\qquad
    \subfloat[][The number of metastasis cases (true positives) and non-metastasis cases (false positives) in the identified subchort using two tests (vs one in L figure), generated using two runs of Algorithm~\ref{algmain}. Sensitivity is proportional to the height of the blue bar, while Specificity is proportional to the ratio between the height of blue and red bars. The sparsity parameter is proportional to the number of genes used.]{\includegraphics[scale=0.4]{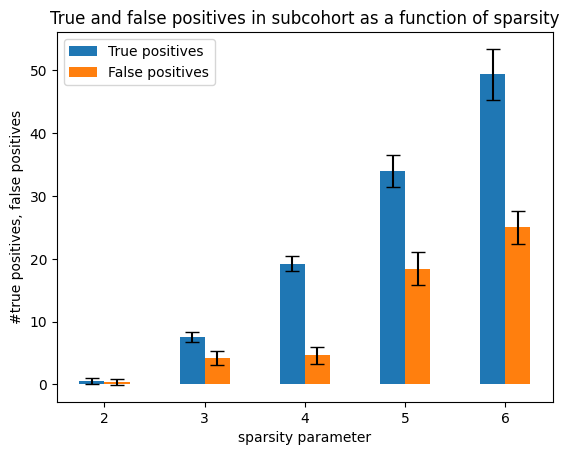}}\\
\caption{Sensitivity vs Specificity trade-off using Algorithm~\ref{algmain}, for the dataset~\cite{metabric}. Left: Algorithm~\ref{algmain} outputs a single test hyperplane that identifies subcohorts with 57\% specificity, for sparsity parameter $s=6$. Right: When we use the intersection of subcohorts using two tests, the specificity increases to about 70\%, but identified subcohort size (sensitivity) decreases to a third, with $s=6$. Note: the single test in the left figure uses about 15 nuclear receptors to identify subcohorts, while the two tests in the right figure use about 45 nuclear receptors.
}
\label{fig:comp1}
\end{figure*}

Figure~\ref{fig:comp1} shows the sensitivity, specificity and sparsity trade-offs for our subcohort identification algorithm. To generate Figure~\ref{fig:comp1}, Algorithm~\ref{algmain} uses the nuclear receptor expressions of a subset of 50 metastatic and 50 non-metastatic cases from the full dataset, the sparsity parameter $s$, a threshold $\theta$, and an arbitrary metastasis patient that is assumed to belong to the subcohort to be identified. The algorithm outputs a test $h$, where $h$ is just a tuple of weights for the nuclear receptors. On a separate hold-out set of a 1000 patients (500 with metastasis and 500 without), the test simply calculates the weighted average value of the given nuclear receptor expression levels for each patient in the hold-out, and if this value is below the threshold then the patient is added to the subcohort, otherwise not. The above exercise is repeated $10$ times for various sparsity values, and sensitivity and specificity measured each time, to arrive at Figure~\ref{fig:comp1}.

{\em Interpretation of Figure~\ref{fig:comp1}:} First, as the number of genes increases, so do specificity and sensitivity in Figure~\ref{fig:comp1}. Second, sensitivity can be traded-off for specificity: We run Algorithm~\ref{algmain} twice, on two different subsampled sets of patients, and obtain two half-spaces/tests $h_1, h_2$ that correspond to two subcohorts. The new subcohort is defined to be the intersection of the previously identified half-spaces, i.e., both tests $h_1,h_2$ have to place patients in the subcohort for the patient to be in the new subcohort. Clearly, specificity should increase for this more conservatively defined subcohort and sensitivity should decrease. Indeed that is the case -- specificity increases to 70\% while sensitivity decreases by a factor of 3 (see Figure~\ref{fig:comp1}(b)). One may even combine three or more tests in this way. Thus one can use Algorithm~\ref{algmain} to increase the specificity of the subcohort at the cost of decreased sensitivity, i.e., decreased size of the identified subcohort.

{\em Comparison:} We compare our algorithm with a version of the well-known PRIM algorithm in Appendix (Section~\ref{sec:prim}). It is clear from our experiments that the specificity of the MAX-CUT based approach outperforms that of the local search based heuristic when sensitivity is held constant. This is despite the fact that the local search heuristic was allowed an order of magnitude more time for computations.

\subsection{Methylation and gene expression}\label{sbs:nn}
\begin{algorithm}[htb]
\caption{Identify Critical Genes}\label{algmain2}
\begin{algorithmic}[1]
\Statex {\bf Input:} Dataset $D\subset \RR^{d}$, and parameter $\epsilon\in(0,1)$ (the $d$ coordinates are the genes)
\Statex {\bf Output:} A subcohort with critical genes that are significantly differently expressed from the population
\Statex
\State Uniformly sample $S\subset D$ with $|S|\simeq\epsilon\cdot|D|$ 
\State Use Algorithm~\ref{algmain} on $S$ to compute a half-space $h$ corresponding to a likely homogeneous subcohort.
\State Compute means and standard deviation of each of $d$ coordinates in population and subcohort.
\State {\bf If} any coordinate has mean which is one standard deviation away from population mean, then {\bf return} the coordinate (gene), subcohort and $h$ (test)
\State {\bf Else} generate a new sample (Step 1) and repeat
\end{algorithmic}
\end{algorithm}
One application of Algorithm~\ref{algmain} is to identify genes which differ significantly between a subcohort and the population. Algorithm~\ref{algmain2} presents the basic idea. We will use a slight variation of it to relate epigenetic changes (nucleotide methylation changes) to clinically significant genetic changes downstream -- an important question in genomics (see for example~\cite{epi1}). 
The use of subsets is critical as averaging over the entire dataset leads to large confidence intervals.

\begin{figure*}[htb!]
    \centering
    \subfloat[][Expressions of nuclear receptors in metastasis and non-metastasis.]{\includegraphics[scale=0.45]{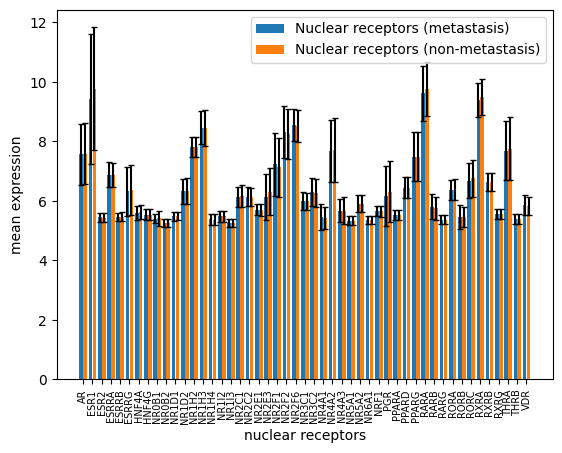}}\qquad
    \subfloat[][Percentage methylations of breast cancer related genes in metastasis and non-metastasis.]{\includegraphics[scale=0.45]{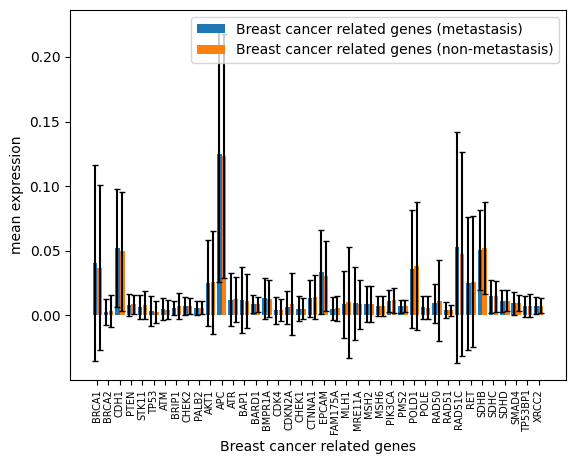}}\\
\caption{Gene expression and methylation percentages for metastasis and non-metastasis patients computed over the full dataset. Observe that averaging over the entire dataset, consisting of non-homogeneous patient population, leads to large confidence intervals. The latter make it impossible to distinguish between metastasis and non-metastasis using z-scores based on the nuclear receptors and breast cancer related genes used above.}
\label{fig:poplevel}
\end{figure*}

{\em Background:} Methylation of certain tumor suppressor genes can potentially lead to breast cancer progression~\cite{epi1,brcgenes}, and nuclear receptors are a target for a large number of drugs~\cite{nrdrugs,nrdrugs2}. Therefore, if we compute the mean methylation and expression levels of critical genes over the entire patient population in the dataset to find a tumor suppressor gene that is significantly more methylated and a nuclear receptor that is significantly over-expressed; then we can test whether a drug targeting that over-expressed nuclear receptor is effective in patients that show a similar pattern of excessive methylation and over-expression. Unfortunately, the situation is not so simple -- there are no such sets of tumor suppressors and nuclear receptors where significant change in one is associated with significant change in the other. Figure~\ref{fig:poplevel} shows that the confidence intervals are just too large, when we average over the entire dataset (because of inhomogeneity in the disease). However, it is possible that the disease state amongst the patients in some subcohorts is homogeneous enough to allow for narrow confidence intervals, and thus identify any simultaneous significant deviation in tumor suppressor methylation and nuclear receptor expression, amongst patients restricted to such a subcohort. Indeed that is the case below.

{\em Algorithm~\ref{algmain2} for relating epigenetics with nuclear receptors:} We use a variant of Algorithm~\ref{algmain2}, with about $1000$ different random subsamples of inputs (of $100$ patients each), from the original dataset (of size $2000$ patients) to generate patient subcohorts of  breast cancer patients (and corresponding tests to identify the subcohorts) in two ways: (1) based on methylation percentage levels, and (2) based on nuclear receptor expression levels. 

Next, for every pair of subcohorts generated (one from methylation levels and another from nuclear receptor expressions), if the two subcohorts have more than a certain percentage of patients in common (we used a threshold of 10\% due to limited dataset size) i.e., they share a relatively large subset of the patient population, then {\em they are likely to be homogeneous as two different sets of markers -- genetic and epigenetic -- are identifying a similar subset of patients}. 

Next, we compute the mean methylation and expression together with confidence intervals, for each gene using the common patients. Finally, we iterate until we find a patient subcohort with genes that have large deviations from the population means.

\begin{figure*}[htb!]
    \centering
    \subfloat[][The nuclear receptors in the identified subcohort, consisting mainly of metastasis cases, that are predicted to differ significantly in expression from their mean population values. In particular, for patients in this subcohort, NR1H2 is over-expressed. Statistically insignificant bars are not shown.]{\includegraphics[scale=0.45]{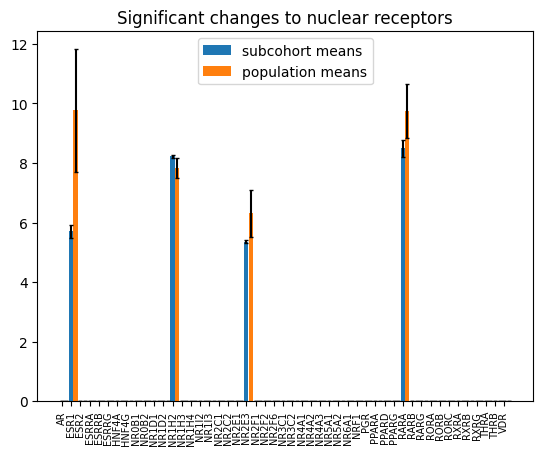}}\qquad
    \subfloat[][The methylation percentages in the identified subcohort, consisting mainly of metastasis cases, that are predicted to differ significantly in expression from their mean population values. In particular, for patients in this subcohort, BRCA2 and MSH6 were methylated more, both are related to DNA repair, and errors can lead to breast cancer progression~\cite{brca2}. Statistically insignificant bars are not shown.]{\includegraphics[scale=0.45]{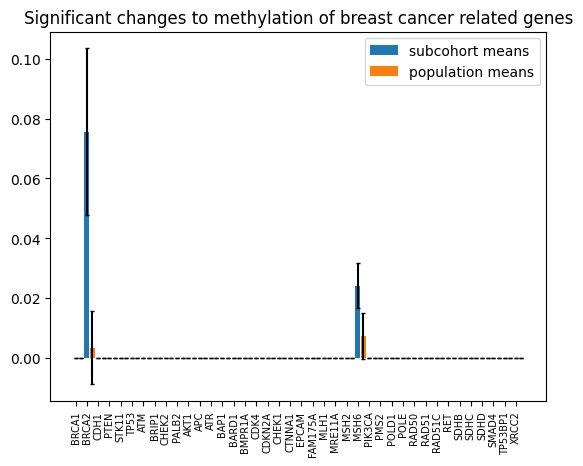}}\qquad
    \subfloat[][The coefficient weights of the genes in the test output by Algorithm~\ref{algmain} to identify the subcohort above.]{\includegraphics[scale=0.45]{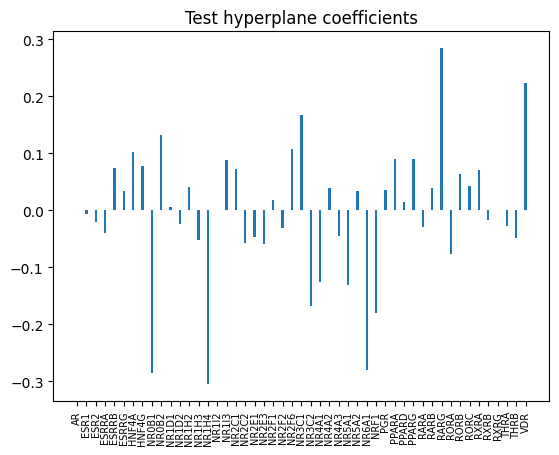}}\qquad
    \subfloat[][The coefficient weights of the genes in the test output by Algorithm~\ref{algmain} to identify the subcohort above.]{\includegraphics[scale=0.45]{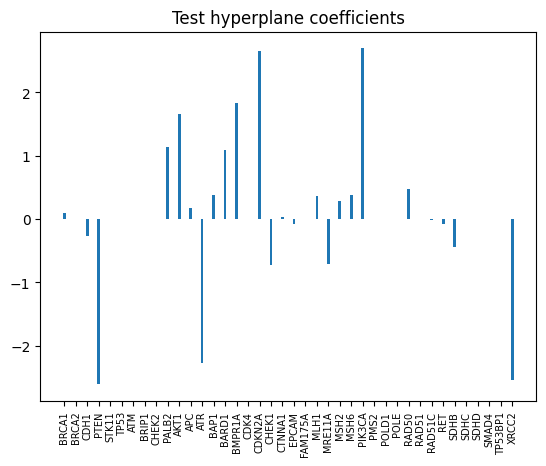}}
\caption{Associating changes in methylation percentages with changes in nuclear receptors using Algorithm~\ref{algmain}. Compared to averaging over the entire population (see Figure~\ref{fig:poplevel}), averaging over patients in a subcohort can lead to statistically significant differences between metastasis and non-metastasis breast cancers. In particular, in the above subcohort of patients, BRCA2 and MSH6 have significantly higher methylation percentages than the entire cohort of the patients in METABRIC dataset; and at the same time they have a significantly higher expression of NR1H2 receptors responsible for lipid metabolism.}
\label{fig:assoc}
\end{figure*}

One such  subcohort is shown in Figure~\ref{fig:assoc}.


{\em Interpretation of Figure~\ref{fig:assoc}:} Patients in the subcohort are identified by sparse linear tests, where the coordinate weights (coefficients) for the test hyperplane are given in Figure~\ref{fig:assoc}(c,d). Interestingly, this particular subcohort of patients shows significant methylation increase in BRCA2 and MSH6. Moreover, nuclear receptor NR1H2 (equivalently LXRB) is significantly over-expressed when compared to the average non-metastatic patient in the dataset. Therefore, LXR-inverse agonists (see for example~\cite{lxr-ant}) may prove to be a useful therapeutic target for this subcohort of patients, if indicated by further clinical evidence. 

\section{Conclusion and limitations}
In conclusion, we have demonstrated a method to design tests that identify homogeneous subgroups (subcohorts) of patients. This was done by translating the problem to a MAX-CUT like formulation (as opposed to a rule-based decision tree like formulation), which admits a nearly optimal efficient approximation algorithm via an extension of the Goemans-Williamson algorithm~\cite{gw}. The latter may be independently interesting. Furthermore, the identified subcohorts and accompanying tests could help broaden the set of patients for existing treatments (for e.g. breast cancer patients and LXR inverse agnosts). However, we do not have the resources to verify our clinical predictions and we hope our algorithm (or similar idea) can be adapted by oncologists to pursue it further.


\clearpage

\clearpage
\bibliographystyle{plain}
\bibliography{main}
\clearpage
\thispagestyle{empty}

\onecolumn
\section{Proof of Theorem~\ref{thm:main}}\label{sec:proof}
\begin{proof}
We need to show that if the SDP solution has value OPT then the rounded solution has value at least $\alpha\mathrm{OPT}$, for some constant $\alpha$ (to be determined). Hence it suffices to upper-bound the ratio of each term in the in the SDP solution objective by the corresponding term in the objective for the expectation of the rounded solution. In other words, we need to upper-bound:
\begin{equation}\label{eqn:r}
r:=\frac{2-2\langle v_u,v_{u'}\rangle-2\langle v_{u_0},v_{u'}\rangle+2\langle v_u,v_{u_0}\rangle}{\mathbb{E}\left[(z_u-z_{u'})^2+(z_{u_0}-z_{u'})^2-(z_u-z_{u_0})^2\right]},
\end{equation}
where $z_u,z_{u'}$ and $z_{u_0}$ are obtained upon rounding $v_u,v_{u'}$ and $v_{u_0}$ using the Gaussian rounding procedure defined in Algorithm~\ref{algmain}. Now imagine the three unit vectors $v_u,v_{u'}$ and $v_{u_0}$ as lying on the unit circle, and let $x\in[0,\pi]$ be the angle between $u_0,u$ and $x'\in[0,\pi]$ be the angle between $u_0,u'$. 

Next, consider each term in the denominator, $\mathbb{E}\left[(z_{u_0}-z_{u'})^2\right]$ equals the probability that the random chosen Gaussian -- that is equivalent to choosing a uniformly random point on the unit circle (with high probability) --  is at an acute angle with $u_0$ and an obtuse angle with $u'$ or vice versa. Thus,
\begin{equation}
\mathbb{E}\left[(z_{u_0}-z_{u'})^2\right] = 4\cdot\frac{x'}{\pi}.    
\end{equation}
Similarly, one has:
\begin{eqnarray}
\mathbb{E}\left[(z_{u_0}-z_{u})^2\right] &=& 4\cdot\frac{x}{\pi}
\end{eqnarray}
Two cases arise:
\begin{eqnarray}
x+x'\le\pi:\ \mathbb{E}\left[(z_{u'}-z_{u})^2\right] &=& 4\cdot\frac{x+x'}{\pi}\\  
x+x'\ge\pi:\ \mathbb{E}\left[(z_{u'}-z_{u})^2\right] &=& 4\cdot\frac{2\pi-x-x'}{\pi}.
\end{eqnarray}
Furthermore, the numerator equals:
\begin{equation}
    2-2\langle v_u,v_{u'}\rangle-2\langle v_{u_0},v_{u'}\rangle+2\langle v_u,v_{u_0}\rangle = 2-2\cos(\theta)-2\cos(x')+2\cos(x),
\end{equation}
where we have used $\cos(\pi-x)=\cos(x)$ and $\theta$ is either $x+x'$ (for $x+x'\le\pi$), or $\theta$ is either $2\pi-x-x'$ (for $x+x'\ge\pi$). Thus the ratio $r$ in Equation~\ref{eqn:r} equals:\footnote{Just to compare, for the well-known rounding in~\cite{gw}, one finds the corresponding ratio to be: $r_{GW}=\frac{1-\cos(x)}{x}$, and the maximum is  well-known to be at $x\simeq 134\mathrm{deg}$.}
\begin{eqnarray}
x+x'\le\pi:\ r(x,x') = \frac{\pi}{4}\cdot\frac{2-2\cos(x+x')-2\cos(x')+2\cos(x)}{2x'}\label{eqn:lepi}\\
x+x'\ge\pi:\ r(x,x') = \frac{\pi}{4}\cdot\frac{2-2\cos(2\pi-x-x')-2\cos(x')+2\cos(x)}{2\pi-2x}.\label{eqn:gepi}
\end{eqnarray}
Note that whether $x'\to0$ in Equation~\ref{eqn:lepi} or $x\to\pi$ in Equation~\ref{eqn:gepi}, the denominator goes to $0$ but the numerator goes to $0$ as well. Moreover, since $\cos(x)$ is even, the numerator goes to $0$ as a quadratic in both cases, so the limit is $0$. Thus $0$ or $\pi$ are likely not the maximum points.

For $x+x'\le\pi$: Taking the partial derivative with respect to $x$, one gets: $\sin(x+x')-\sin(x)=0$. Thus either (1) $x+x'=x$, so $x'=0$; or (2) $x+x'=\pi-x$, so $2x=\pi-x'$. However, we have ruled out (1). Substituting (2) into Equation~\ref{eqn:lepi}, we get (at the extremum):
\begin{eqnarray}
r(x) &=& \frac{\pi}{4}\cdot\frac{1-\cos(x+x')-\cos(x')+\cos(x)}{x'}\\
&=& \frac{\pi}{4}\cdot\frac{1-\cos(\pi/2+x'/2)+\cos(\pi/2-x'/2)-\cos(x')}{2x'}.
\end{eqnarray}
Plotting the single variable function above, shows that the maximum value of $r$ is $1.22$ and is reached at $x=1.86$rads.

For $x+x'\ge\pi$: Taking the partial derivative with respect to $x'$ and equating with $0$ shows: $\sin(x')-\sin(x+x'-\pi)=0$ at the extremum. Thus either (1) $x'=x+x'-\pi$, so $x=\pi$; or (2) $\pi-x'= x+x'-\pi$, so $x'=\pi-x/2$. However, we have ruled out (1). Substituting with (2) into Equation~\ref{eqn:gepi}, we get (at the extremum):
\begin{equation}
    r(x) = \frac{\pi}{4}\cdot\frac{1-\cos(\pi-x/2)-\cos(\pi-x/2)+\cos(x)}{\pi-x} = \frac{\pi}{4}\cdot\frac{1-2\cos(\pi-x/2)+\cos(x)}{\pi-x}.
\end{equation}
Plotting the single variable function above, shows that the maximum value of $r$ is $1.22$ and is reached at $x=1.28$rads.

Thus the maximum value of $r$ is $1.22$, and the rounded solution is within $\frac{1}{1.22}=0.82$ of the optimal SDP solution (and hence the optimal solution). Finally, note that the ratio of vertex set sizes, i.e.,$\frac{|S_U|}{|S_U^*|}$, equals the ratio of cut sizes as long as $|S_U^*|\ll |U|$. Hence the proof follows.
\end{proof}

\section{Proof of Theorem~\ref{thm:spec}}\label{sec:app-spec}
\begin{proof}
Without loss of generality assume that the variable $z_{u_0}$ in Equation~\ref{eqn:obj1} is $+1$ (if not one can flip all the signs). Next, note that each term of the objective of the optimization problem in Equation~\ref{eqn:obj1} is either $+1$ or $-1$ (for the $\pm 1$ solution). The total value of the cut is assumed to be $\kappa\cdot|U||V|$. Therefore, if $x$ and $y$ denote the fraction of $U$ and $V$ vertices on the subcohort side of the cut, then we have by comparing the leading term in the objective with the cut value above:
\begin{equation}
2x(1-y)\cdot|U||V|\ge \kappa\cdot |U||V|
\end{equation}
However, the cut we compute using SDP rounding is not optimal but only close to it, i.e., 
\begin{equation}\label{eqn:spec-constr2}
    2x(1-y)\cdot|U||V|\ge \alpha\kappa\cdot |U||V|,
\end{equation} 
with $\alpha\simeq0.82$ (cf. Theorem~\ref{thm:main}). 

Finally, the specificity is the ratio of $U$ points on selected side of the cut with all $U$ and $V$ points on the subcohort side of the cut, i.e., specificity is given by the ratio of $U$ vertices over all vertices on the subcohort side of the cut, i.e.,
\begin{equation}
    \frac{x|U|}{y|V|+x|U|}
\end{equation}
In order to lower bound the ratio, we simply need to compute its minimum over $x,y\in(0,1)$ under the constraint in Equation~\ref{eqn:spec-constr2}. Substituting $y$ with $1-y$ proves the result.
\end{proof}

\begin{figure}[htb!]
    \centering
    \subfloat[][Heatmap depicting lower bound on specificity as a function of $\kappa$ (x-axis) and $\frac{|V|}{|U|}$ (y-axis)]{\includegraphics[scale=0.45]{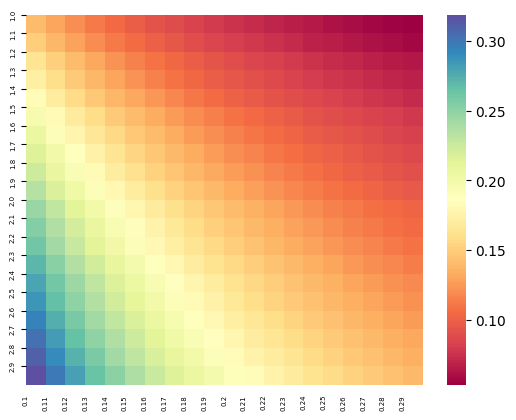}}\qquad
    \subfloat[][Lowerbound on specificity vs $\kappa$ (max. possible sensitivity)]{\includegraphics[scale=0.42]{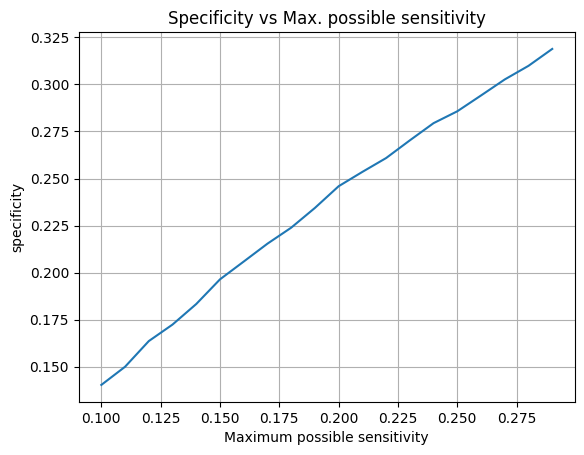}}
\caption{Lower bound on specificity as a function of $\kappa$ and $\frac{|V|}{|U|}$, based on Theorem~\ref{thm:spec}.}
\label{fig:spec_surf}
\end{figure}

One may think of $\kappa$ as the maximum fraction of edges that can be in the cut given the sparsity constraints. Thus $\kappa$ is the upper bound on sensitivity, while the value of the optimum in Equation~\ref{eqn:spec-opt} is a lower bound on the sensitivity. Thus we plot the objective in Equation~\ref{eqn:spec-opt} in Figure~\ref{fig:spec_surf} for various values of $\kappa$ and $\frac{|V|}{|U|}$, to illustrate the lower bound on specificity.
\section{Proof of Theorem~\ref{thm:constr}}
\begin{proof}
A straight forward application of Lemma~\ref{lm:grot2} (also known as Grothendieck's identity) implies that the RHS of the constraints in Equations~\ref{eqn:rlx1} and~\ref{eqn:rlx2} equals $-\frac{2\theta}{\pi}$ and $\left(1-\frac{2}{\pi}(z_{u'}z_{u_0})\right)$ respectively, in expectation. Thus the constraints of our SDP relaxation are not violated excessively. Hence the proof follows.
\end{proof}

\begin{lemma}\label{lm:grot2}
Given a $2$-dimensional Gaussian $(X,Y)$ with mean zero and covariance matrix $\begin{bmatrix} \delta_x & \rho \\ \rho & \delta_y \end{bmatrix}$, we have:
    \begin{equation}
        \mathbb{E}[\mathrm{sign}(X)\mathrm{sign}(Y)] = 1-2\mathbb{P}\left(\frac{Z_1}{Z_2}\le t\right) = \frac{2}{\pi}\arcsin\left(\frac{\rho}{\sqrt{\delta_x\delta_y}}\right),
    \end{equation}
where $Z_1$ and $Z_2$ are independent mean zero Gaussians with covariance $\rho$ and variance $1+\frac{\rho^2}{\delta_x\delta_y}$ and $\delta_y$ respectively.
\end{lemma}
\begin{proof}
\noindent Note that, $Z:=\frac{X}{\sqrt{\delta_x}}-\frac{\rho Y}{\sqrt{\delta_x}\delta_y}$ is a mean zero Gaussian independent of $Y$, and has variance $1+\frac{\rho^2}{\delta_x\delta_y}$.

\begin{eqnarray}
    \PR\left(XY\ge 0\right) &=& \PR\left(ZY\ge \frac{\rho}{\sqrt{\delta_x}\delta_y}Y^2\right)\\
    &=& \PR\left(\frac{Z}{\sqrt{1+\frac{\rho^2}{\delta_x\delta_y}}}\ge\frac{\rho Y}{\sqrt{\delta_x}\delta_y\sqrt{1+\frac{\rho^2}{\delta_x\delta_y}}}\right)\\
    &=&\PR\left(\frac{\tilde{Z}}{\tilde{Y}}\ge\frac{\rho}{\sqrt{\delta_x\delta_y}\sqrt{1+\frac{\rho^2}{\delta_x\delta_y}}}\right),
\end{eqnarray}
where $\tilde{Y}$ and $\tilde{Z}$ are independent standard Gaussians, and thus their ratio is the Cauchy distribution. Therefore, the last probability above can be calculated as:
\begin{eqnarray}
    \PR\left(XY\ge 0\right) &=& \frac{1}{\pi}\arctan\left(\frac{\rho}{\sqrt{\delta_x\delta_y}\sqrt{1+\frac{\rho^2}{\delta_x\delta_y}}}\right)\\
    &=& \frac{1}{\pi}\arcsin\left(\frac{\rho}{\sqrt{\delta_x\delta_y}}\right).
\end{eqnarray}
This completes the proof of Lemma~\ref{lm:grot2}.
\end{proof}
\section{Comparison with PRIM (continued from Subsection~\ref{sbs:pred})}\label{sec:prim}
In this section, we compare our algorithm's performance against local search based algorithms (PRIM based subgroup discovery).

As a baseline, we implemented a version of the PRIM subgroup discovery algorithm~\cite{ff}. PRIM is a local search algorithm, where the subgroups are defined by axis-parallel hyper-rectangles. The algorithm work as follows: One starts with a hyper-rectangle covering the entire dataset, and peels off a small portion of the rectangle from one or more dimensions (sides) in each iteration of the algorithm. In each iteration, one computes and stores the specificity for the subgroup defined by the points within the rectangle and after trying a sufficient number of rectangles one returns the optimal subgroup (with highest specificity) seen thus far. In this way the algorithm gradually trades-off sensitivity for increased specificity.

In our case, we additionally want to ensure a sparsity constraint, and then compute the specificity for a fixed sensitivity in order for comparison with our algorithm. So we make the following modifications:
\begin{itemize}
    \item We enforce our sparsity constraint by using rectangles of dimension at most $k$ that are chosen as random subsets of size $k$ from $\{1,...,50\}$. The reason for using $50$ is that in our case the subsets correspond to subsets of the 50 possible nuclear receptors.
    \item We require that the generated rectangle have at least $\sigma$ fraction of points that correspond to patients with metastatic disease ($\sigma$ was chosen to be either 0.1 or 0.3 for comparison). Thus ensuring a minimal sensitivity for the generated subgroups.
\end{itemize}

Finally, we capped our search by generating 400K axis-parallel rectangles. It required about $5\times$ the computational time needed compared to Algorithm~\ref{algmain}.

\begin{figure*}[htb!]
    \centering
  \includegraphics[scale=0.45]{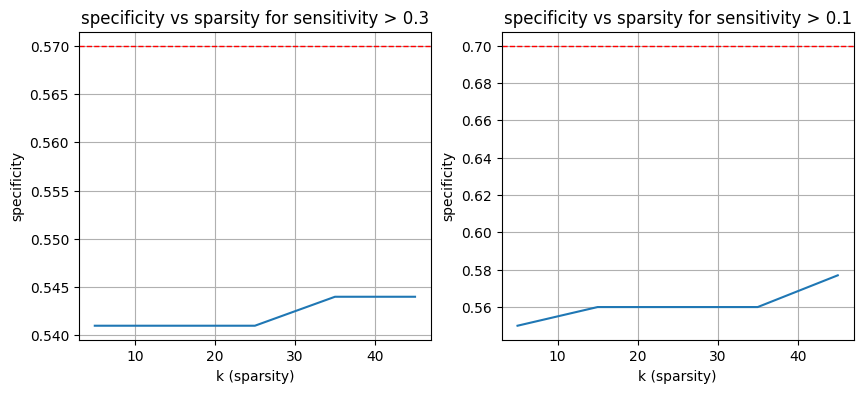}
\caption{Plot specificity vs k (sparsity) as sensitivity is held above a lower bound for the variant of PRIM algorithm used for comparison. The dashed Red line (in both figures) denotes the specificity for the MAX-CUT algorithm, from Figure~\ref{fig:comp1}, for the given sensitivities ($0.3$ and $0.1$), when the number of genes used is 15 (L figure) and 45 (R figure). In both cases, the PRIM local search heuristic fails to approach the specificity of Algorithm~\ref{algmain}.}
\label{fig:prim_comp}
\end{figure*}

{\em Baseline (PRIM) vs Experiment (MAX-CUT) specificity:} The results for the rule-based local search PRIM algorithm are shown in Figure~\ref{fig:prim_comp}.


Therefore, the MAX-CUT based linear separator approach shows better specificity than the rule-based PRIM algorithm for a given sensitivity. This is not unexpected as Algorithm~\ref{algmain} is not a simple local search confined to an axis-parallel hyper-rectangle, but a more sophisticated SDP rounding based algorithm that computes a general (non-axis-aligned) sparse linear separator.


\end{document}